# Comparison of Canonical Correlation and Partial Least Squares analyses of simulated and empirical data.

## Anthony R McIntosh[1,2,3]

[1]Dept of Biomedical Physiology & Kinesiology, Simon Fraser Univ
[2]Rotman Research Institute, Baycrest Centre
[3]Dept of Psychology, University of Toronto

Email: randy_mcintosh@sfu.ca
Mailing Address:
Biomedical Physiology and Kinesiology,
Faculty of Science
Simon Fraser University
8888 University Drive Burnaby, BC V5A 1S6
Canada

## Abstract

**Background**

With the availability of large datasets containing multiple measures, there has been a renewed interest in applying multivariate statistical analysis. Two methods, Canonical Correlation Analysis (CCA) and Partial Least Squares (PLS), have been used most frequently, given their historical links to classic statistical modelling of the dimensions that relate to two data blocks. Though similar in the decomposition of the cross-block structure, there are important differences in specific steps. In this paper, we apply the most general form of CCA and PLS to three simulated and two empirical datasets, all having large sample sizes on the order of n=10,000. We take successively smaller subsamples of these data to evaluate sensitivity, reliability, and reproducibility.

**Results**

In null data having no correlation within or between blocks, both methods showed equivalent false-positive rates regardless of sample size. Both methods also showed comparable detection of effects in data with weak but reliable effects until sample sizes drop below n=50. In the case of strong effects, both methods showed similar performance unless the correlations of variables within one data block were high. In these instances, the reproducibility in CCA declined. This was ameliorated if a principal components analysis (PCA) was performed on a data block, and the component scores were used to calculate the cross-block matrix. This solution is only valid if the PCA structure is stable with respect to the population. For PLS, the results were reproducible across sample sizes for strong and moderate cross-block effects, regardless of the within-block correlations, but show lower detectability at small sample sizes (n=20).

**Conclusions**

The general outcome of our examination gives three messages. First, for data with low within and high between block structure, CCA and PLS give comparable results, with equivalent sensitivity and false-positive rate. Second, if there are high correlations within either block, this can compromise the reliability of CCA results. This can be remedied with PCA before cross-block calculation. However, this assumes that the latent variable structure extracted with PCA is stable for a given sample. Third, the statistical significance by null hypothesis testing does not guarantee that the results are reproducible, even with large sample sizes. This outcome suggests that researchers should routinely assess both statistical significance and reproducibility of their data.

**Keywords**: multivariate statistics, reproducibility, data analytics, resampling, bootstrap

Background

With the trend in collecting large datasets, there has been a parallel increase in the interest in multivariate statistical methods to extract the complicated patterns that relate data features. In the neuroimaging domain, data sets such as the UKBioBank, Imagene and Enigma, and Human Connectome Project provide a formidable challenge in the diversity of data spanning genetic, demographic, and various phenotypic measures and neuroimaging data on brain structure and function. One goal for these initiatives is to identify the dimensions in one data set that predict or explain dimensions in the other - for example, how does brain structure relate to lifestyle and cognitive measures? While univariate methods can be used for a general first impression, they are typically underpowered and cannot take advantage of the added sensitivity of extracting combinations of variables that link data sets[1, 2]. A distributed pattern of brain activity may be correlated with a battery of neuropsychological measures but not easily discernible when each brain area or neuropsychological measure is considered alone. These combinations of variables are generally referred to as Latent Variables (LV), reflecting the fact that they represent a dimension in the data that is inferred from a collection of measures (e.g., 'memory' is inferred from a collection of memory tests; the default mode network is inferred from a composite of core regions).

The first multivariate method focused on the extraction of LVs relating two data blocks is canonical correlation analysis (CCA). Introduced by Hotelling [3], the technique extracts LVs that maximize the correlation between data blocks X and Y. Each LV contains weights for variables within block that indicate the relative contribution of each to the LV. In the most basic form, the LVs are mutually orthogonal, with each accounting for successively less of the correlations between data sets. The weights for the variables within each LV are adjusted for the mutual intra-block correlations, thus indicating the unique contribution each variable makes to a given LV. In this way, CCA minimizes redundancy across variables within blocks

A more recent method called Partial Least Squares (PLS) has a similar goal to CCA in deriving LVs that reflect the relationship between two data blocks. The original version of the method was developed by Wold[4], with links to multi-table forms of Tucker[5].  There have been two popular versions, PLS-R and PLS-C, where PLS-R is focused primarily on directed predictions of X on Y, while PLS-C is non-directed and most similar to CCA [6]. Thus we will concentrate on PLS-C for this paper. PLS-C operates by optimizing the covariance between data tables, extracting mutually orthogonal LVs that account for maximum covariance. The weights for the variable within each LV reflect the contribution of the variable to the LV *without* adjusting for the mutual intra-block correlations, which is an essential difference from CCA.

CCA and PLS have been applied to neuroimaging datasets[6-10]. Some attempts have compared the performance of the two methods with some uncertainty [11, 12]. Part of the inconsistency seems to stem from how the two methods are implemented, whether the data examples are empirical or simulated, and the assumptions for the simulations. Our goal here is to focus on the most general forms of CCA and PLS-C, using empirical and simulated data to characterize where the two methods agree and where they depart. The code for all analyses is also open, allowing the interested reader to evaluate our results and extend them.

We start by reviewing the theory behind both methods We begin with a null data set with no reliable cross-block relations and then two empirical datasets with differing magnitudes of cross-block associations. We finish with another simulated dataset with many variables in one block, which is meant to capture the case in Big Data sets such as neuroimaging or genetics. This final dataset provides clear evidence of divergence and convergence of PLS and CCA. These examples support four conclusions that are elaborated in the discussion. The details of the methods for analyzing and assessing significance and reliability are described after the discussion.

# Theory

As noted above, CCA is one of the most general forms of multivariate analysis and is the core of the General Linear Model (GLM)[1]. Let us assume there are two data blocks in matrices *X* and *Y*, where both conform to a multivariate normal distribution. The rows of these matrices have *n* observations and columns of *p* variables for *X* and *q* variables for *Y*.

The calculation for the correlation matrices:

1) $R_{XX} = (X^t * X) * (n-1)^{-1}$
2) $R_{yy} = (Y^t * Y) * (n-1)^{-1}$

Where matrices *X* and *Y* are centred and scaled by their column-wise standard deviation (i.e., z-score transform). Superscript "t" is matrix transpose, and "-1" is the matrix inverse, and '*' is matrix multiplication.

The cross-block correlation matrices *Rxy* and *Ryx*:

3) $R_{xy} = (X^t * Y) * (n-1)^{-1}$
4) $R_{yx} = (Y^t * X) * (n-1)^{-1}$

One implicit assumption is that *X* and *Y* contain some underlying relationships among variables such that the correlation matrices *Rxx* and *Ryy* are not identity. The second more explicit assumption is that the cross-block correlation *Rxy* exists (has non-zero elements), and is the focus of the statistical test.

For CCA, before the decomposition of the cross-block matrix is done, it is adjusted for the within-block correlations:

5) $\Omega = Rxx^{-0.5t} * Rxy * Ryy^{-0.5t}$

Most often, $\Omega$ is used for singular value decomposition (SVD):

6) $[U, S, V] = SVD(\Omega)$

Where

7) $\Omega = U * S * V^t$

and

8) $U^t * U = I, V^t * V = I$

Where "$I$" indicates an identity matrix.

The diagonal matrix $S$ contains singular values that are the canonical correlations for each LV. Matrices U and V contain the singular vectors corresponding to the weights for each variable in X and Y, respectively. The first columns of each have the weights for the first LV, with the corresponding first element of S having the canonical correlation for that LV. Subsequent LVs will have lower canonical correlations, accounting for increasingly less of $\Omega$.  (Note that $\Omega$ is sometimes calculated as: $Rxx^{-1t} * Rxy * Ryy^{-1t} * Ryx$. In this case, the singular values of the decomposition are the squared canonical correlation)

For PLS, the main difference is that the decomposition is performed on $Rxy$ directly without adjustment:

9) $[U, S, V] = SVD(Rxy)$

In this case, the singular values in matrix $S$ are the covariances rather than the correlation, even though $Rxy$ may contain correlations. This is because the scaling of $Rxy$ for CCA is equivalent to scaling that happens in the conversion of covariance to a correlation.

To illustrate, consider if $X$ and $Y$ here are vectors that are mean-centred only, the covariance Cxy is:

10) $Cxy = X^t * Y * (n-1)^{-1}$

And respective covariances of X and Y alone (i.e., variance)

11) $Cxx = X^t * X * (n-1)^{-1}$
12) $Cyy = Y^t * Y * (n-1)^{-1}$

And the correlation $Rxy$

13) $Rxy = Cxx^{-0.5} * Cxy * Cyy^{-0.5}$

This is similar to the calculation done for $\Omega$ in equation(5). Hence, a correlation coefficient is a scaled covariance.

This reinforces the point that CCA optimizes the correlation between $X$ and $Y$, whereas PLS optimizes the covariance because of the differences in scaling the cross-block matrix rather than the contents within.

A second, and more important point, is that the cross-block scaling limits the data structure that CCA can handle. If either $Rxx$ or $Ryy$ are rank deficient, the calculation in equation(5) is impossible. In a less severe case, when there are high correlations among variables within $X$ or $Y$, the scaling in equation (5) will alter $Rxy$ so that the CCA and PLS results will diverge. If there are high correlations within-block, one can calculate the *canonical structure coefficients*, which are the zero-order correlation of each item within-block with the canonical variate. This is done by calculating the standardized canonical weights and then rescaling by the within-block correlation. For the $X$ matrix, the standardized canonical weights, $A$, would be:

14) $A = Rxx^{-0.5} *$

And the canonical structure coefficients $S_A$:

15) $S_A = Rxx * A$

When there are high within-block correlations, $A$ and $Sa$ will diverge; hence, both are commonly presented for CCA publications[13]. It is also the case that $S_A$ will be more similar to PLS weights in $U$ for the same $Rxy$ matrix.

In CCA, the challenge of high within-block correlations can also be addressed by performing a dimensionality reduction with Principal Components Analysis (PCA) on one or both data blocks first and using the component scores as input data. The component scores are mutually orthogonal; thus, $Rxx$ and/or $Ryy$ would be identity (I). The operation in equation (5) effectively divides $Rxy$ by one (an identity matrix), so the differences between CCA and PLS are reduced or eliminated. The use of PCA/ICA as a dimensionality reduction step before statistical testing has been a long practice [14-17], so it will be evaluated here.

The issue with high within-block correlations for CCA is ubiquitous with GLM approaches to statistical analyses where parameter estimation for a given variable is typically adjusted to reflect their unique contribution over and above the variance shared with other variables. Whether this is the desired outcome depends on the investigator's theoretical orientation. As we shall see below, the issue can also affect the reliability of parameters in CCA.

Foreward: In what follows, we compare PLS and CCA with four metrics (see Methods for the details) .

1) Null hypothesis tests to determine whether the parameter estimates could come from random data. This was done using permutation tests and conventional parametric test (Bartletts test).
2) Detection of Effects. In the case where the null hypothesis was rejected, how many significant effects (LVs) were detected
3) Reliability: are the parameters estimates reliable after repeated resampling?
4) Reproducibility: are the parameter estimates reproducible in repeated train-test samples?

These metrics are compared as a function of sample size within and between PLS and CCA.

## Data Set 1 - Null data

As a simple assessment of the false positive rate for PLS and CCA, we generated a data set with no reliable correlations among any variables using the *randn()* function in Matlab. The function takes the number of rows and columns for the output and draws from a normal distribution with a mean of zero and standard deviation of 1. We generated 10,000 observations with 10 X variables and 5 Y variables.  As shown in Figure 1, the within and between block correlations are weak, all hovering close to zero.

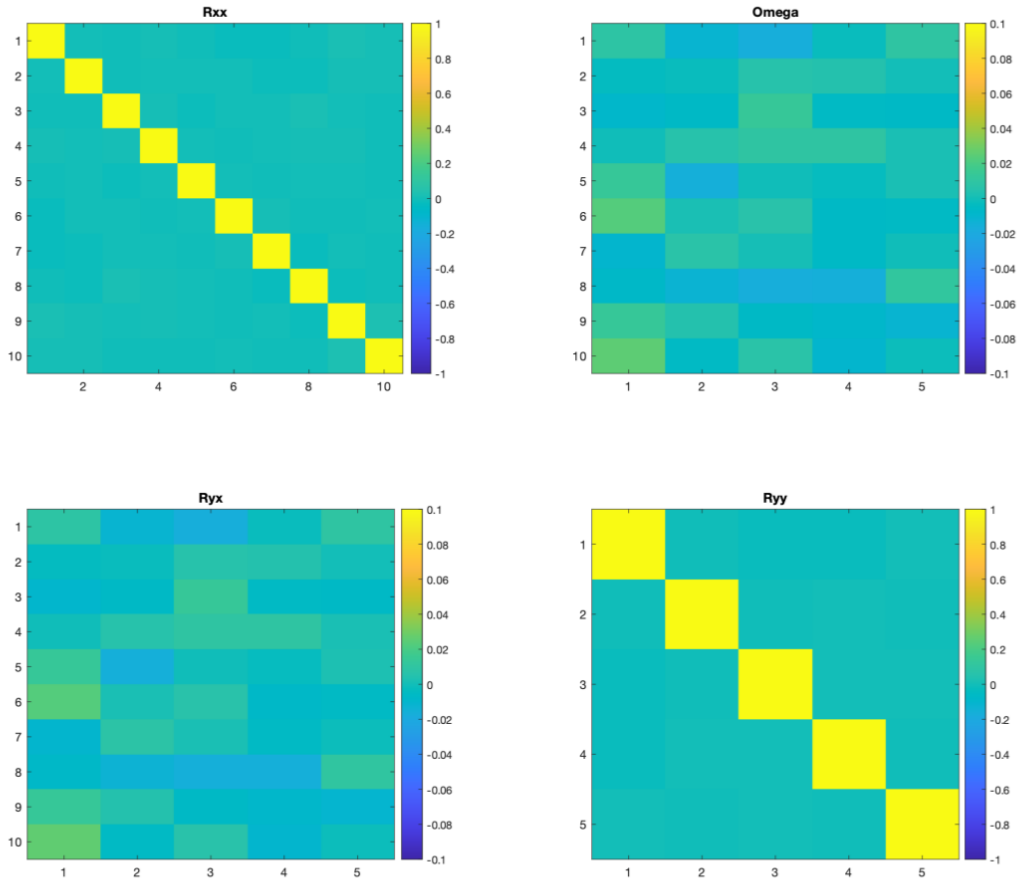

*Figure 1. Within (Rxx, Ryy) and between (Ryx, Omega) block correlations for data set 1. Colour scale (indicated by colour bar) is set differently for within and between block correlations.*

CCA and PLS performed on these data identified no significant LVs by permutation test, which was confirmed by Bartlett's test for CCA (see Table 1).

*Table 1.  Singular values and permutation probability values for each Latent Variable (LV) in dataset 1. Bartlett's test result for CCA is in italics below the table.*

|  | PLS | CCA |
|---|---|---|
| LV1 | 0.046, p = 0.532 | 0.045, p = 0.570 |
| LV2 | 0.039, p = 0.286 | 0.039, p = 0.260 |
| LV3 | 0.022, p = 0.946 | 0.022, p = 0.943 |
| LV4 | 0.021, p = 0.479 | 0.021, p = 0.472 |

| LV5 | 0.009, p = 0.835 | 0.009, p = 0.826 |

Bartlett's test for all LVs; $\chi^2(50)$ = 46.47, P ~0.62

As expected, the reproducibility assessment indicated that none of the LVs were reproducible across multiple split-half samples (Table 2). For train/test assessment, the z-score for the test samples hovered around zero and for singular vectors less than 1.95 (roughly 95% confidence interval).

*Table 2 - Reproducibility metrics*
*z-scores for train/test on singular values*

|     | PLS | CCA |
| --- | --- | --- |
| LV1 | -0.02 | -0.06 |
| LV2 | 0.03 | 0.049 |
| LV3 | -0.26 | -0.25 |
| LV4 | -0.32 | -0.32 |
| LV5 | -0.255 | -0.24 |

*z-scores for singular vectors from split-half*

|     | PLS | CCA |
| --- | --- | --- |
| LV1 | 1.32, 1.51 | 1.34, 1.47 |
| LV2 | 1.50, 1.57 | 1.50, 1.57 |
| LV3 | 1.42, 1.55 | 1.41, 1.55 |
| LV4 | 1.40, 1.59 | 1.41, 1.59 |
| LV5 | 1.45, 1.57 | 1.43, 1.57 |

Finally, we ran 500 iterations of subsampled data, ranging from N=500 to N=20, conducting PLS and CCA for each iteration. Permutation tests were performed for each iteration, and once complete, we calculated the number of iterations where a p-value for any LV was <= 0.05. This yielded an estimated false-positive rate that averaged 0.05 for all sample sizes (mean: 0.052; range: 0.046-0.056). The main conclusion from this simulation is that CCA and PLS show a similar false-positive rate across a broad range of sample sizes when using resampling statistics for null hypothesis testing.

# Empirical data - COSMO

Next are the results from two analyses of a relatively large sample of survey data from the COSMO initiative in Germany[18]. The details of the data collection protocol can be found here: https://www.psycharchives.org/handle/20.500.12034/2392. The data from the first 13 waves were analyzed. Two portions of these data are considered. For the first, we looked at weak but robust effects that were detectable in the full sample. Here the emphasis was on how the detectability of small effects changed with progressively smaller samples.  In the second, we looked at a sub-sample focusing on highly correlated indicators that also showed strong cross-block relations. Here the focus was both on detectability, and the reproducibility of the parameter estimates, given the high correlations among the variables.

## COSMO Set 1

The total sample across the 13 waves was 11,669. The Y block consisted of age (mean 46yrs, range 18-87), education split into low (9 years or less, n=1183), medium (10 yrs without university, n=3961) and high (10 yrs. with university education, n=6525), and sex (M=5842, F=5827). Education was left as a single vector rather than split into two indicators to account for the additional degree of freedom (e.g., dummy coding). Since we are not interpreting the analytic outcome per se, we left education as a single vector. The analytic outcome below did not change if education was left as a single or double indicator.

The X block consisted of 6 Affect measures that indicated the degree a participant perceived the threat of COVID. These were Distance (COVID infection is near me), Fear (I am fearful of COVID), Hype (The threat of COVID is too hyped), Spread (COVID is spreading quickly), Think (I think about COVID often), and Worry (I am worried about COVID).  Participants indicated their responses on a 7-point scale.

The within and cross-block correlations are shown in Figure 2. The scales for Ryx and Omega show that the values for cross-block correlations are small (-0.15 to 0.2). The within block correlations for Y are similarly weak, but the values are larger for X, going as high as 0.64. The high correlations in X meant that the cross-block correlations differ between Ryx and Omega, which is most evident in the correlations of X indicators with the first column of Y (age).

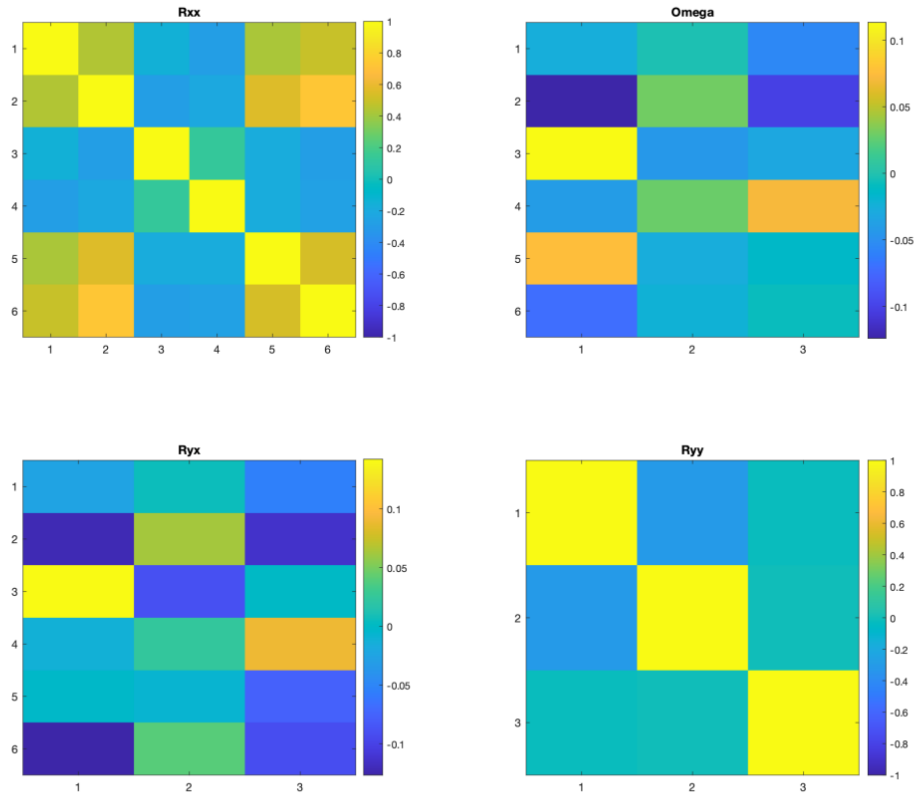

*Figure 2. Within (Rxx, Ryy) and between (Ryx, Omega) block correlations for COSMO data set 1. Colour scale (indicated by colour bar) is set differently for within and between block correlations.*

Permutation tests were equivalent for CCA and PLS, as was Bartlett's test for CCA (Table 3).

*Table 3 - Singular values and permutation probability values for each Latent Variable (LV) in COSMO data set 1. Bartlett's test results for CCA are below*

|  | PLS | CCA |
|---|---|---|
| LV1 | 0.286 p=0 | 0.214  p=0 |
| LV2 | 0.149 p=00 | 0.135 p=0 |
| LV3 | 0.019 p=0.004 | 0.04 p=0 |

*Bartlett's Test: All LVs: $\chi^2(18) = 781.09$. P~0.0; LVs2-3; $\chi^2(10) = 234.82$. P~0.0; LV3 $\chi^2(4) = 19.74$, P~0.005*

The singular values for PLS and CCA are similar, suggesting that the cross-block adjustment did not have an appreciable effect on the cross-block correlations in Omega (Figure 2).  The third LV has a small canonical correlation (0.04) yet was significant, emphasizing that a large sample size can attribute significance to small effects. The next question is whether all three of these LVs show equivalent reproducibility.

*Table 4 - Reproducibility metrics*
*z-scores for train/test on singular values*

|     | PLS   | CCA   |
| --- | ----- | ----- |
| LV1 | 18.4  | 23.37 |
| LV2 | 14.66 | 14.81 |
| LV3 | 1.47  | 3.2   |

*z-scores for singular vectors from split-half (X, Y)*

|     | PLS        | CCA         |
| --- | ---------- | ----------- |
| LV1 | 81.3, 67.8 | 46.02, 56.92 |
| LV2 | 50.1, 57.4 | 28.23, 40.54 |
| LV3 | 1.89, 81.96 | 3.93, 53.26 |

Table 4 shows that for both PLS and CCA, the first two latent variables show high reproducibility but diverge at the third. The overall train/test for LV3 in PLS would not be considered reproducible with a z=1.47, while CCA has a high value of z=3.2.  A similar trend is noted for the singular vectors, particularly with vectors for the X matrix, which shows less reproducibility for PLS (1.89 vs. 3.93).  The Y matrix has a higher z-score, likely due to its lower dimensionality relative to X.

The subsample tests confirmed the general observation that weak effects are detected only with very large sample sizes. The bar graph in Figure 3 shows the detectability for the three LVs by sample size. For both CCA and PLS, there is an immediate drop in detectability with N=500, though the first two LVs are detected about 75% of the time. This drops to around 50% for N=250, 25% for N=100, 10% for N=50 and down close to 5% for N=20.

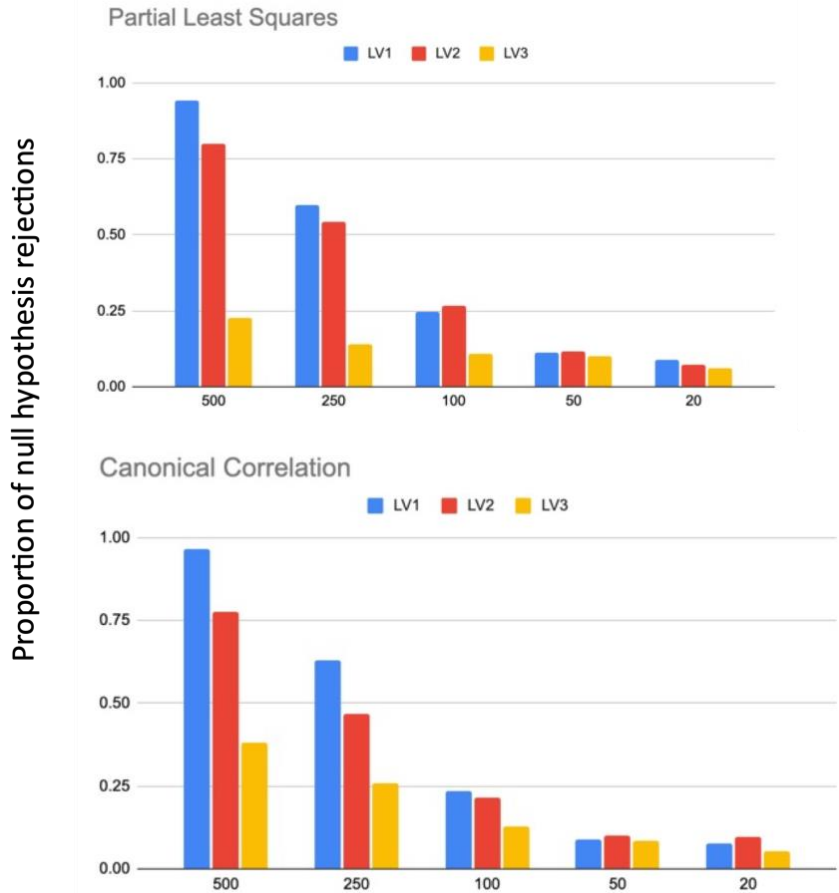

*Figure 3. Detectability by Sample Size. . The bars indicate the proportion the null hypothesis was rejected for each LV for PLS and CCA.*

The reproducibility assessment by sample size mirrored what was seen for the detectability test, where the reproducibility declined with sample size both for singular values and singular vectors.  We focused on the first LV for the singular values (Table 5), showing the reproducibility for LV1 was weak for N=500 and essentially absent for smaller samples sized.

*Table 5. Average z-score train/test singular value and split-half for singular vectors for X and Y in parentheses from the first latent variable*

|  | PLS | CCA |
|---|---|---|
| 500 | 3.35 (5.17, 4.17) | 2.90 (3.4, 4.11) |
| 250 | 2.03 (3.64, 2.90) | 1.54 (2.23, 2.61) |
| 100 | 0.89 (2.65, 2.28) | 0.63 (1,71, 2.00) |
| 50 | 0.48 (2.33, 2.11) | 0.27 (1.58, 1.85) |
| 20 | 0.20 (2.15, 1.97) | 0.03 (1.51, 1.72) |

The split-half test for the singular vectors mirrored the singular value test with decent reproducibility for the first LV for N=500 and 250, but dropping with smaller sample sizes whereby N=100 the singular vectors are poorly reproduced across split-halves.  The decline was essentially the same for PLS and CCA.

## COSMO Set 2.

We extracted a subset from the COSMO data above to assess the situation where the within-block correlations were strong. We used data from 4 of 13 waves (4, 5, 8, 13), where subjective mental health was measured. Matrix X contained six resilience items with alternative valence (Recover Quickly from Hard Times, Difficult to Endure Stressful Events, Recover Fast from Stress, Hard to Get to Usual Self After Unpleasant Event, Manage Difficult Time without Difficulty, Take Long Time to Recover from Setback). Matrix Y contained four measures of psychological well-being (Nervous, Depressed, Lonely, Hopeful).  All measures ranged from 1 to 5. The total sample size was 3605, and the analysis was collapsed across age and sex.

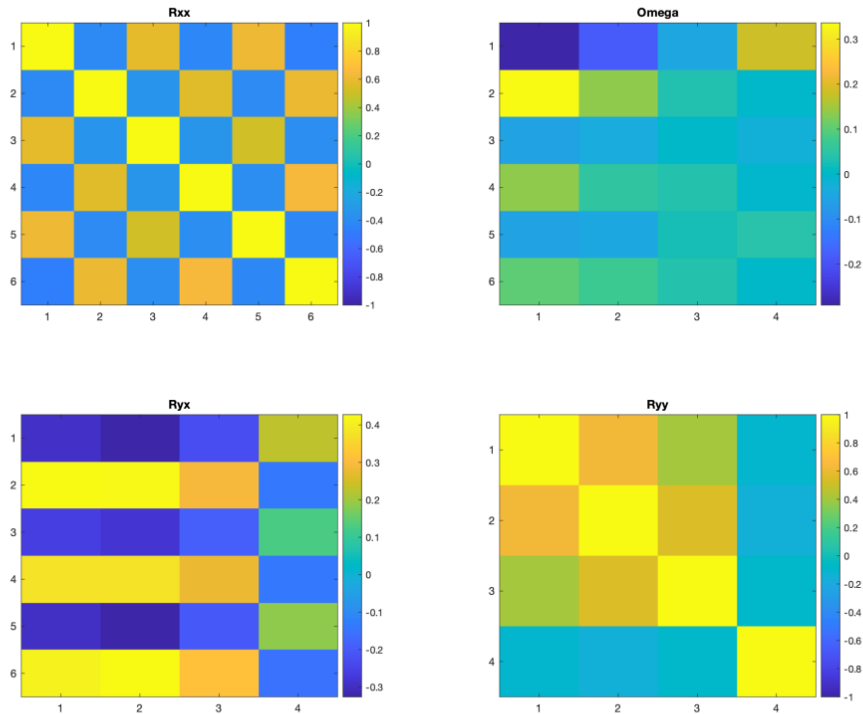

*Figure 4. Within (Rxx, Ryy) and between (Ryx, Omega) block correlations for COSMO data set 1. Colour scale (indicated by colour bar) is set differently for within and between block correlations.*

Unlike the previous COSMO dataset (Figure 2), the within and between block correlations were substantially higher (Figure 4).  For the X block, the resilience measures show opposing positive and negative correlations because of the valence of the questions (e.g., greater vs. less resilience). The Y block too had higher correlations, especially among the first three indicators. The high within block correlations resulted in a substantial difference in the patterns for Ryx and Omega. The adjustment for Omega concentrated most of the cross-block relation to the first two indicators in X.

The null hypothesis test diverged for CCA and PLS (Table 6). The permutation test for PLS suggested two LVs were significant, while CCA suggested four. Interestingly, Bartlett's test for CCA also suggested that only two LVs were significant.

*Table 6. Singular values and permutation probability values for each Latent Variable (LV) in COSMO data set 2. Bartlett's test results for CCA are below*

|          | PLS             | CCA               |
|----------|-----------------|-------------------|
| LV1      | 1.438 p=0       | 0.5617  p=0       |
| LV2      | 0.123 p=0       | 0.1476 p=0        |
| LV3      | 0.0251 p=0.108  | 0.0484 p=0.004    |
| LV4      | 0.0121 p = 0.205 | 0.0319 p=0.006   |

*Bartlett's Test: All LVs: $\chi^2(24) = 1455.5$. P~0.0; LVs2-4; $\chi^2(15) = 91.42$. P~0.0; LV3-4 $\chi^2(8) = 12.27$, P~0.14, LV4 $\chi^2(3) = 4.14$, P~0.25*

Despite the differences in the significance test, both methods converged on the reproducibility assessment, suggesting that only the first two LVs were reproducible at both the singular values and the singular vectors (Table 7).

*Table 7 - Reproducibility metrics*
*z-scores for train/test on singular values*

|          | PLS   | CCA   |
|----------|-------|-------|
| LV1      | 37.53 | 46.64 |
| LV2      | 7.14  | 7.38  |
| LV3      | 0.58  | 0.44  |
| LV4      | 0.44  | 0.70  |

*z-scores for singular vectors from split-half (U, V)*

|      | PLS          | CCA          |
|------|--------------|--------------|
| LV1  | 933.5, 663.3 | 194.1, 247.1 |
| LV2  | 19.3, 35.32  | 12.08, 16.62 |
| LV3  | 1.52, 2.34   | 1.48, 1.66   |
| LV4  | 1.58, 2.36   | 1.66, 1.66   |

The detectability tests as a function of sample size showed that the PLS and CCA are similar for the first LV up until sample sizes of 50 and 20 (Figure 5). For PLS, detectability for the first LV stays relatively high across all samples, but the second LV drops rapidly to about 15% for n=100. CCA shows a more gradual decrease in detectability except at N=20, where all LVs sit roughly at 15%. This may be due to the high within block correlations that add instability to parameter estimates more in CCA than PLS.

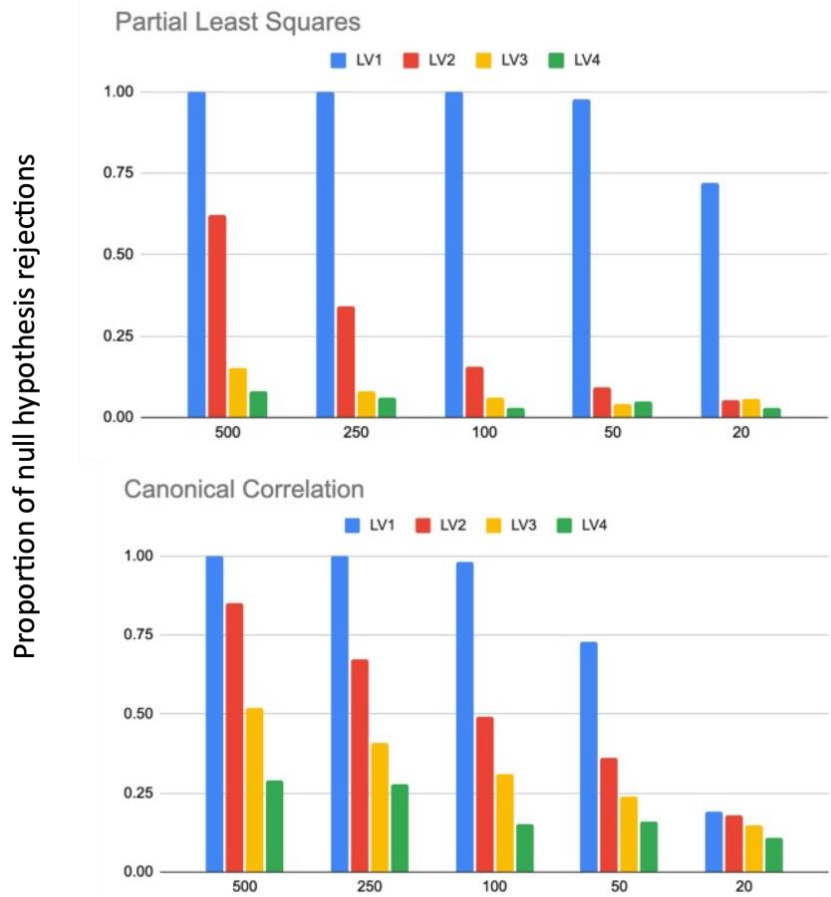

Figure 5. Detectability by Sample Size. The bars indicate the proportion the null hypothesis was rejected for each LV for PLS and CCA.

*Table 8. Average z-score train/test singular value and split-half for singular vectors for X and Y in parentheses from the first latent variable*

|  | PLS | CCA |
|---|---|---|
| 500 | 13.79 (134.53, 93.42) | 15.63 (27.55, 35.35) |
| 250 | 9.85  (68.74, 47.53 | 9.71 (17.11, 13.68) |
| 100 | 6.40 (29.05, 19.95) | 3.73 (4.72, 5.53) |
| 50 | 4.33 (14.38, 9.87) | 1.46 (2.21, 2.46) |
| 20 | 2.15 (4.67, 3.61) | 0.31 (1.31, 1.38) |

The train/test and split-half test for singular vectors (Table 8) also indicated reproducibility declined more rapidly for CCA than PLS. By sample sizes of 50 and 20, the singular vectors for all LVs were not reproducible across all LVs from CCA. Though the results are not shown, the trend for CCA was the same when canonical structure coefficients were used. For PLS, the vectors for LV1 were reproducible across all sample sizes.

# Data Set 4 - Simulated data with a large number of Indicators (SIMREL)

We used SIMREL Field [19] for the final data set to generate an X block with 50 indicators that could be summarized in a lower-dimensional subspace using PCA. This mimics much of the current applications of CCA and PLS to data from Big Data, such as neuroimaging, where a large collection of regions of interest or, in some cases, entire images are used in one data block. Because neuroimaging data have inherent correlations, the data are usually lower-dimensional than the number of indicators (e.g., regions of interest or voxels). This would also mean the X block would have high correlations between indicators. As we reported for COSMO data set 2, the high within-block block correlations affect parameter estimations in CCA more so than in PLS. This effect was explored further in the SIMREL data. One strategy to reduce the impact of high within block correlations is conducting a PCA and using the component scores as the indicators in the X block. This will be explored after we first examine the analysis of the full X and Y blocks.

The SIMREL data were generated such that X and Y blocks were related to two significant dimensions (see Table 8). For the first dimension, 15 X and two Y variables were the dominant contributors. On the second, a different set of 10 X variables and the other two Y variables were dominant. The $R^2$ value for the two dimensions was 0.2 and 0.1, respectively. SIMREL also specifies the components in X related to Y. We indicated that the first and second components were dominant for the first cross-block dimensions and components 3, 4, and 6 for the second cross-block dimension. This specification helped set the expectation for the analysis using PCA on the X matrix before calculating the cross-block matrix.

*Table 9 - SIMREL parameters in R (# indicates comments)*

```
>  simrel(
    +  (n = 10000,      #number of observations
    +  p = 50,                        #number of predictors (X)
    +  q = c(15,10),                  #number of relevant predictors per cross-block component
    +  relpos = list(c(1, 2), c(3, 4, 6)), #X PC's relevant to prediction of Y
    +  gamma = 0.6,                   #decline factor of eigenvalues for X
    +  m = 4,                         #number of responses (Y)
    +  ypos = list(c(1, 3), c(2, 4))),  #relative position of Y on cross-block components
    +  eta = 0,                       #decay factor of eigenvalues for Y
    +  R2 = c(0.2, 0.1),              #population R-squared for cross-block components
    +  type = 'multivariate',         #type of distributions
       )
```

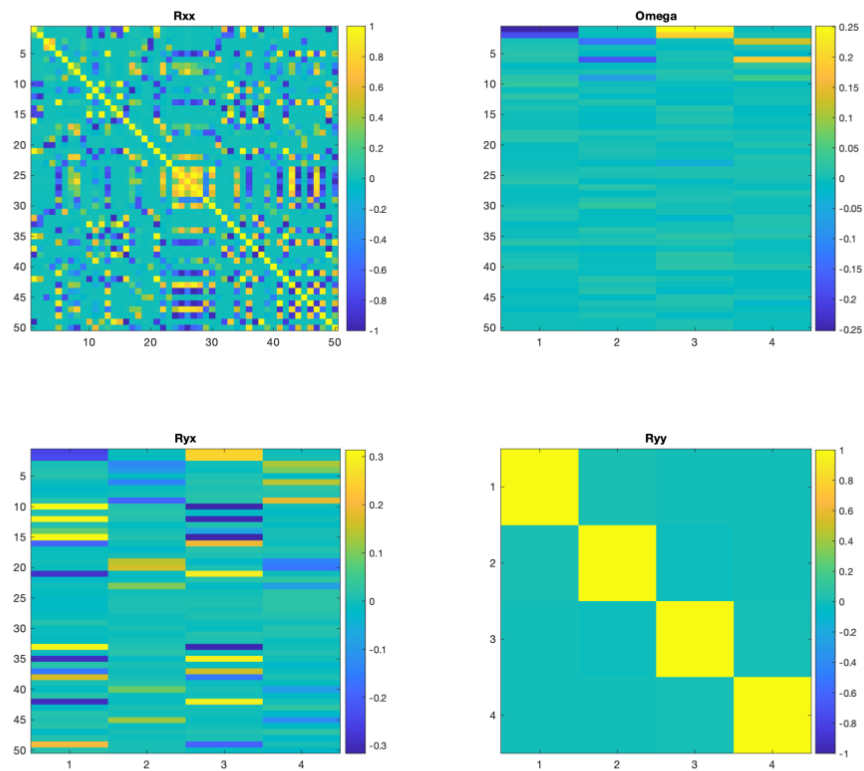

*Figure 6. Within (Rxx, Ryy) and between (Ryx, Omega) block correlations for SIMREL data set 1. Colour scale (indicated by colour bar) is set differently for within and between block correlations*

As with the previous example (see Fig 5), there is a large difference between Ryx and Omega due to the high correlations with the X block. The correlations in the Y block were uniformly low, with a maximum value of 0.01. The pattern of correlations in the X block varied with a tight cluster of indicator numbers 24-30. These indicators were not part of the two cross-block dimensions specified in Table 8 but did affect the CCA and PLS results.

As with the second COSMO data set, the null hypothesis test diverged for CCA and PLS (Table 9). The permutation test for PLS and CCA suggested that three LVs were significant. Interestingly, Bartlett's test for CCA indicated that only two LVs were significant, consistent with how the simulation was constructed.

*Table 10 Singular values and permutation probability values for each Latent Variable (LV) in The SIMREL dataset. Bartlett's test results for CCA are below.*

|  | PLS | CCA |
|---|---|---|
| LV1 | 1.345 p=0 | 0.4517 p=0 |
| LV2 | 0.576 p=0 | 0.3378 p=0 |
| LV3 | 0.094 p=0.001 | 0.077 p=0.004 |
| LV4 | 0.0264 p = 0.71 | 0.062 p=0.09 |

*Bartlett's Test: All LVs: $\chi^2(200) = 3583.5$ P~0.0; LVs2-4; $\chi^2(147) = 1309$, P~0.0; LV3-4 $\chi^2(96) = 100.1$, P~0.36, LV4 $\chi^2(47) = 40.23$, P~0.75*

The reproducibility assessment was the same for PLS and CCA and reinforced the emerging observation that reproducibility and reliability assessment is an essential complement for null hypothesis testing (Table 10). For both the singular values and singular vectors, two LVs were suggested to be reproducible with very large z-scores. At the same time, the last two had very low z-scores, suggesting poor reproducibility.

*Table 11 - Reproducibility metrics*
*z-score for train/test*

|  | PLS | CCA |
|---|---|---|
| LV1 | 46.9 | 55.41 |
| LV2 | 31.09 | 36.8 |
| LV3 | 0.96 | 0.24 |
| LV4 | -0.65 | 0.16 |

*z-score for split-half distribution (X, Y)*

|  | PLS | CCA |
|---|---|---|
| LV1 | 347.03, 308.74 | 91.89, 125.82 |
| LV2 | 71.57, 213.76 | 57.21, 116.01 |
| LV3 | 2.39, 2.48 | 1.34, 2.20 |
| LV4 | 1.68, 2.48 | 1.31, 2.20 |

For this analysis, we also examined the bootstrap ratios of the singular vectors to understand the weights' reliability better and whether the patterns were similar between PLS and CCA. The cosine between the singular vectors for the X block from PLS and CCA was low.  For LV1, the cosine=0.40, and for LV2 cosine=0.53.  This contrasted with the analyses from previous data sets where the cosines were ~0.85 or higher.  The low cosines likely came from the high within-block correlation for X, which resulted in a difference between Rxy and Omega and the derived singular vectors.

Figure 7 shows the distribution of singular vector weights on each LV, with weights having a stable 95% bootstrap confidence interval highlighted. While the V vectors (Y-matrix weights) patterns are identical for PLS and CCA, there were large differences in U (X-matrix weights). For PLS, several indicators show stable parameters in U, while in CCA, a small number of variable weights dominate the first and second LVs.

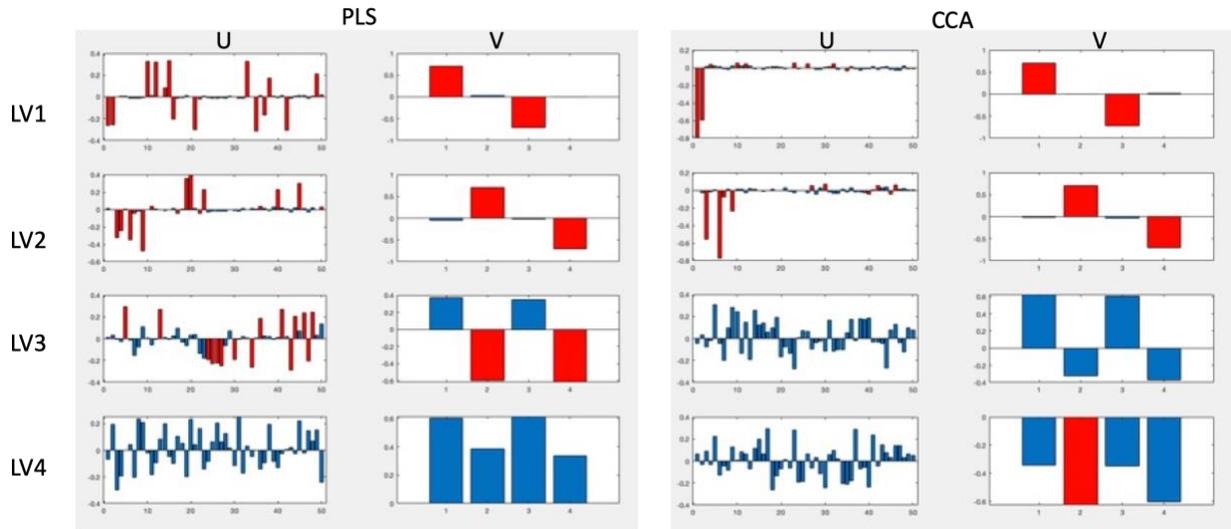

*Figure 7. Singular vector weights for PLS and CCA with parameters highlighted in red if the bootstrap confidence interval did not contain zero.*

A second difference between PLS and CCA is that PLS shows stable weights on the 3rd LV, but CCA does not.

As noted in the theory section, canonical structure weights can be computed to complement the canonical weights. For the present analysis, the canonical structure coefficients are essentially identical to the vector weights for PLSs, with cosines for the first and second LVs at 0.99. In addition, the pattern of reliable structure loadings by bootstrap was similar to PLS for the X variables on the first two LVs (Figure 8).

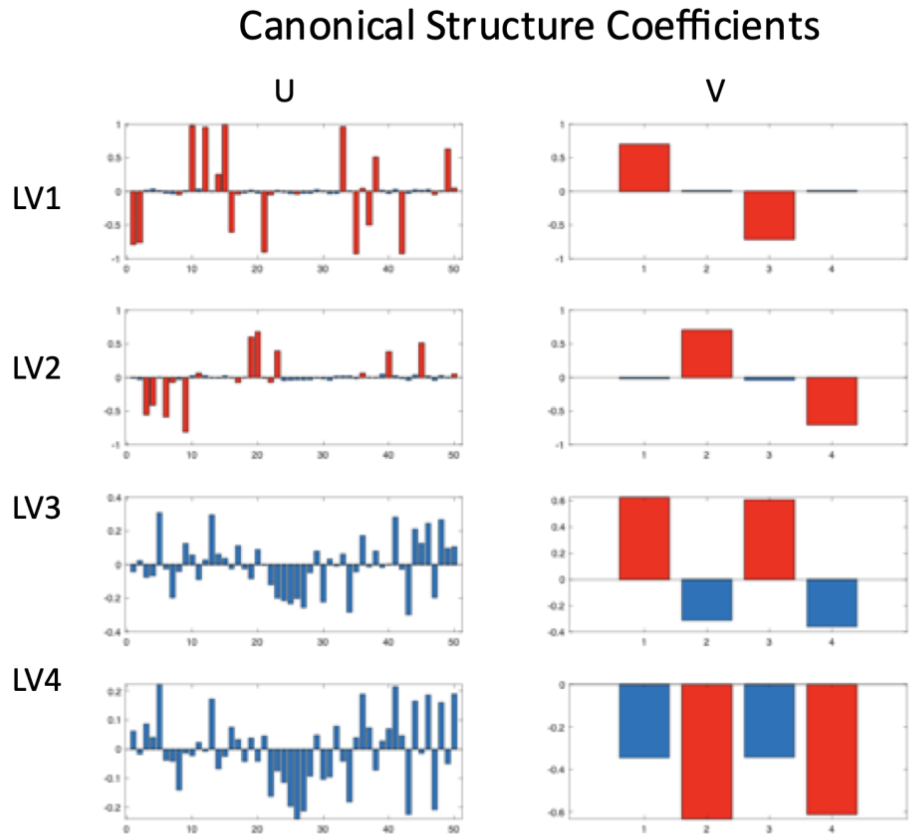

*Figure 8. Canonical structure coefficients. Weights are highlighted in red if the bootstrap confidence interval did not contain zero.*

While using canonical structure coefficients helped overcome the effect of high within block correlations in X, making the outcome similar to PLS, the high correlations also added instabilities with smaller sample sizes, affecting CCA more so than PLS (Figure 9). In the subsampling, we could not assess N=50 and N=20 for CCA as the X matrix was rank deficient. This was also true for the split-half reproducibility test for N=100.

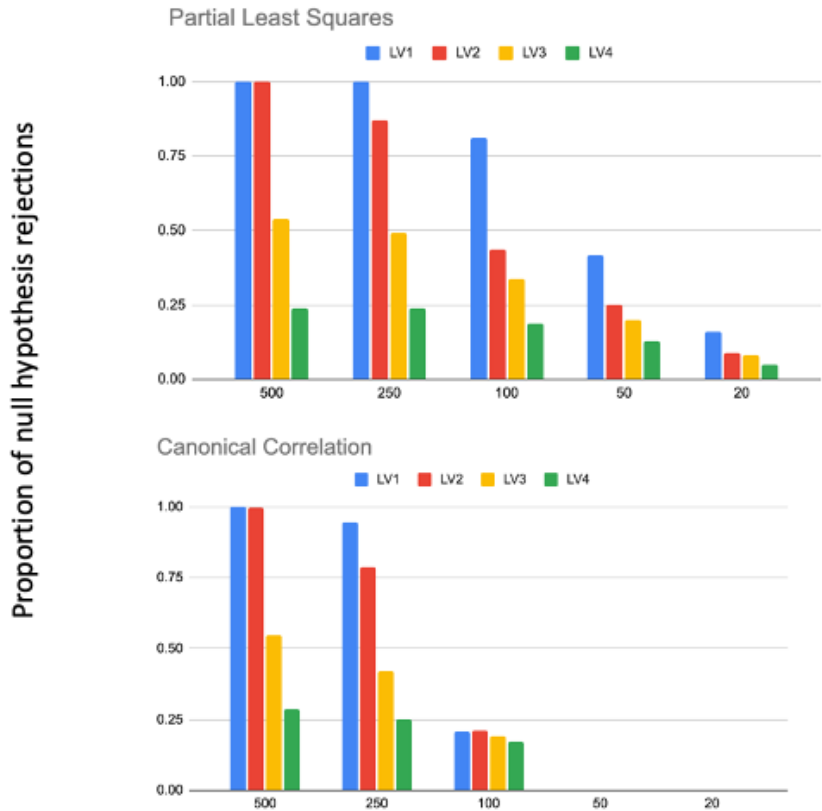

*Figure 9. Detectability by Sample Size. The bars indicate the proportion the null hypothesis was rejected for each LV for PLS and CCA. This was not possible for N=50 and N=20 in CCA because of matrix rank-deficiency.*

In the detectability tests, both methods perform similarly for N=500 and N=250.  They diverged at N=100 where detectability was ~0.1 for all LVs in CCA but remained high for LV1 and LV2 in PLS (Table 11).

Table 12 - Average z-score train/test singular value and split-half for singular vectors for X and Y in parentheses from the first latent variable

|  | PLS | CCA |
|---|---|---|
| 500 | 9.43 (16.86, 15.23) | 5.84 (3.99, 4.29) |
| 250 | 5.67 (8.14, 7.42) | 2.23 (2.10, 2.00) |
| 100 | 2.39 (3.31, 3.13) | NA |
| 50 | 1.13 (2.19, 2.09) | NA |
| 20 | 0.43 (1.83, 1.81) | NA |

The reproducibility tests confirmed the added instabilities in CCA with smaller sample sizes, where the singular value reproducibility, with much lower values for N=500 and N=250, compared to PLS. This reproducibility drop was also noted for singular vectors. For CCA, the reproducibility was around 2.0 at N=250. For PLS, the singular value and singular vectors dropped at N=100, though the vectors for LV1 remained close to 3.0.

## Data Set 4 - Simulated Data with pre-analysis PCA.

As noted in the theory section, one strategy often adopted for large data sets, especially where n<p, is to do a PCA on the larger matrix first and analyze the component scores from a subset of components. We did this for the SIMREL data, taking the first seven components and replacing matrix X with the component scores. Seven components accounted for 98% of the variance in matrix X. This ensured the first six components related to Y were retained, and another component was added to approximate what might be done in an actual analysis. Adding more components did not alter the results.

One challenge for this exercise was to ensure the PCA applied to the sub-samples had a similar pattern of eigenvectors to the PCA for the full sample. While typical data analyses do not have the luxury for such an assessment, we wanted first to ensure the eigenvectors derived from the subsampled data had the same signs as the full population. Thus, for each iteration with each subsample, we corrected for reflections of the eigenvector with respect to the full sample. We did not correct for rotations, assuming that the order of the PCA would be the same as in the full sample.

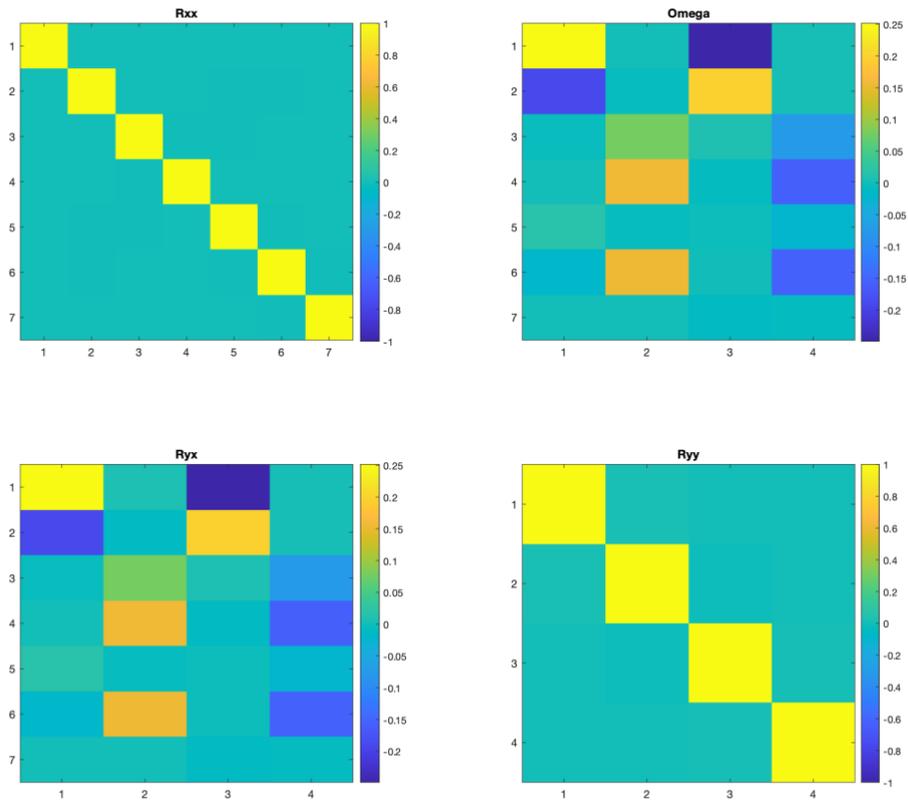

*Figure 10. Within (Rxx, Ryy) and between (Ryx, Omega) block correlations for SIMREL data set using PCA scores for X. Colour scale (indicated by colour bar) is set differently for within and between block correlations*

As would be expected, the within block correlations matrices are identity (or very close to in the case of Ryy), so Ryx and Omega are effectively the same (Figure 10).

The null hypothesis tests for these data were nearly identical for PLS and CCA, with the permutation test indicating that LV1 and LV2 were significant, while LV3 did not pass the conventional threshold. Bartlett's test also suggested that only 2LVs were significant (Table 12).

*Table 13. Singular values and permutation probability values for each Latent Variable (LV) in The SIMREL dataset. Bartlett's test results for CCA are below*

|  | PLS (s and p-value) | CCA (s and p-value) |
|---|---|---|
| LV1 | 0.4478 p=0 | 0.4480  p=0 |
| LV2 | 0.3306 p=0 | 0.3321 p=0 |
| LV3 | 0.024 p=0.09 | 0.0246 p=0.09 |
| LV4 | 0.013 p = 0.30 | 0.0129 p=0.31 |

Bartlett's Test: *All LVs: $\chi^2(28) = 3407.7$ P~0.0; LVs2-4; $\chi^2(18) = 1169$, P~0.0; LV3-4 $\chi^2(10) = 7.73$, P~0.65, LV4 $\chi^2(4) = 1.94$, P~0.75*

The reproducibility assessment was identical for PLS and CCA, indicating that the LV1 and LV2 were reproducible at the singular value and singular vector levels (Table 13).

*Table 14 - Reproducibility metrics*
  *z-scores - train/test*

| | PLS | CCA |
|---|---|---|
| LV1 | 47.52 | 56.93 |
| LV2 | 31.37 | 34.99 |
| LV3 | -0.35 | -0.31 |
| LV4 | -0.11 | -0.12 |

*z-scores from split-half (X, Y)*

| | PLS | CCA |
|---|---|---|
| LV1 | 83.15, 83.60 | 130.25, 131.34 |
| LV2 | 76.27, 80.32 | 111.01, 120.96 |
| LV3 | 1.44, 2.07 | 1.43, 1.47 |
| LV4 | 1.48, 2.07 | 1.47, 2.10 |

The equivalence between PLS and PCA persisted for the bootstrap estimation for the singular vector weights (Figure 11).

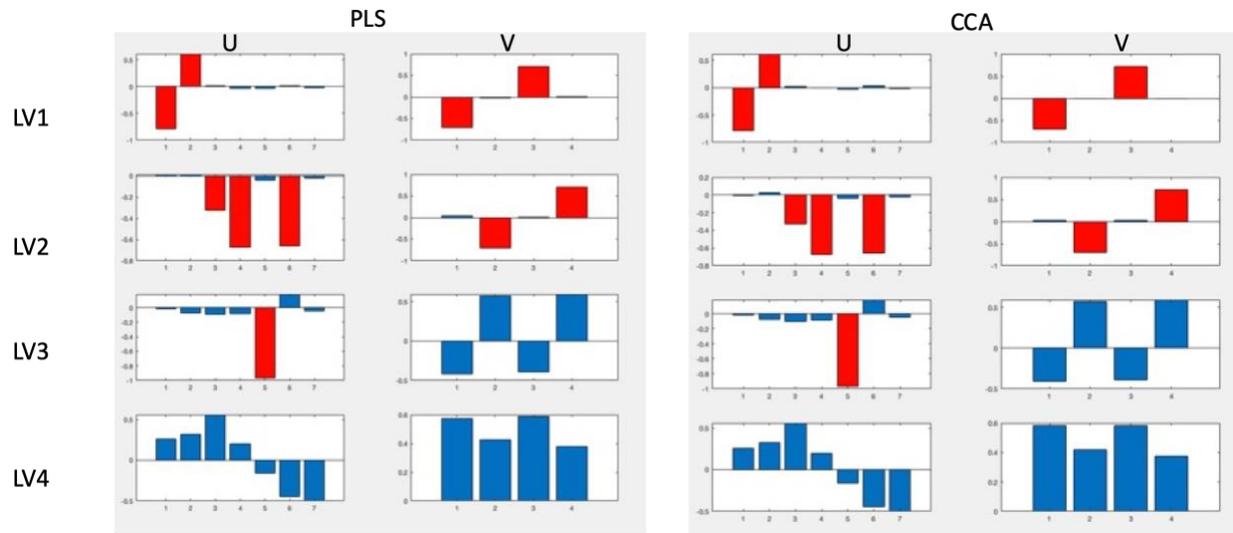

*Figure 11. Singular vectors with stable weights are shaded red.*

This equivalence continued with the subsamples for detectability and reproducibility. PLS and CCA showed the same decline in detectability with sample size, falling to less than 50% for LV1 and LV2 at N=50 (Figure 12).

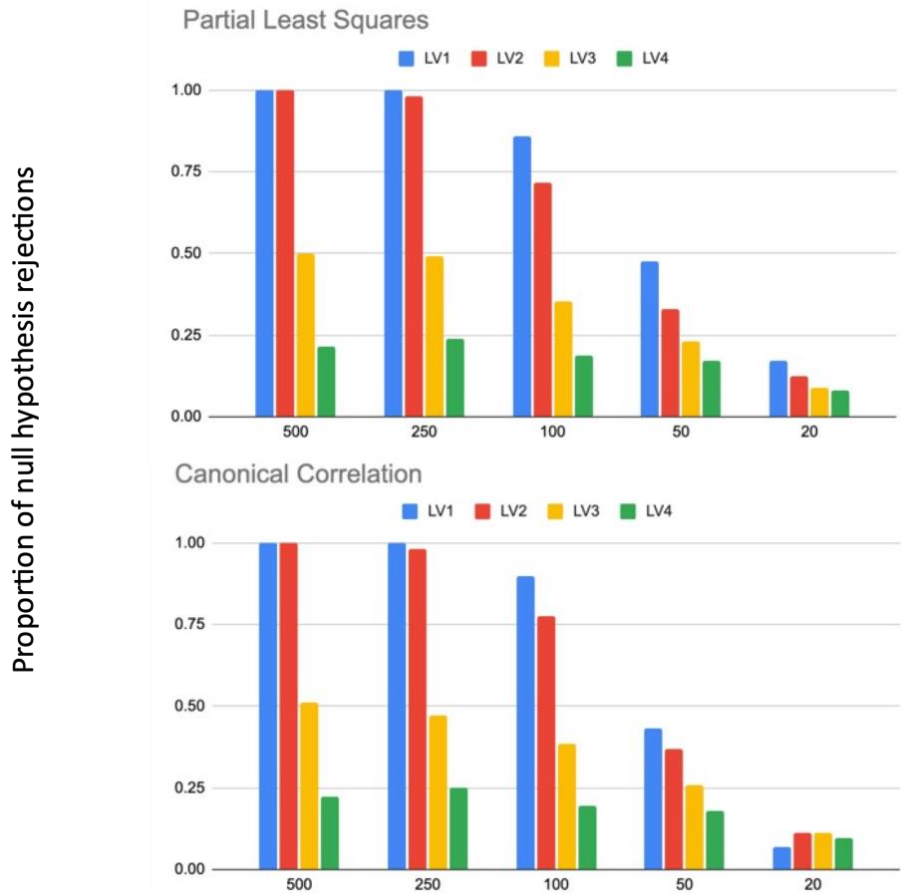

*Figure 12. Detectability by Sample Size. The bars indicate the proportion the null hypothesis was rejected for each LV for PLS and CCA.*

*Table 15 - Average z-score train/test singular value and split-half for singular vectors for X and Y in parentheses from the first latent variable*

|  | PLS | CCA |
|---|---|---|
| 500 | 8.18 (5.93, 6.24) | 9.86 (7.53, 7.79) |
| 250 | 4.94 (3.7, 3.94) | 5.78 (4.59, 4.43) |
| 100 | 2.13 (2.26, 2.45) | 2.3 (2.35, 2.49) |
| 50 | 1.01 (1.77, 1.93) | 0.99 (1.69, 1.81) |
| 20 | 0.34 (1.60, 1,74) | 0.20 (1.48, 1.53) |

Reproducibility for the singular value for the first LV declined roughly the same for PLS and CCA, being just slightly higher than 2.0 for N=100 and then dropping to one or lower. The singular vector reproducibility followed the same trend by sample size.  The parallel trend was also noted for the singular vectors (Table 14), where the reproducibility dropped below 2.0 with sample sizes of 100 and less for both methods.

## Discussion

The purpose of this paper was to make a direct comparison of PLS-C and CCA to determine where they converge and diverge in terms of overall statistical outcome. To do this, we tried to make the comparisons as equivalent as possible by:

1) Using cross-block correlations for both analyses
2) Using resampling statistics as the primary outcome measure
3) Using reliability/reproducibility as a complement to #2
4) Comparing the singular vector weights rather than the derived canonical structure coefficients.

In addition, we examined how sample size affected the detectability of known effects, in the case of simulated data, and of empirical effects of different strengths. Below we enumerate and briefly discuss the primary outcomes

1) ***When the within-block correlations are low, PLS-C and CCA have essentially the same outcome.***

This was shown in the Null Data, the first COSMO data set, and the SIMREL data with pre-analysis PCA. There is nothing surprising about this outcome since the cross-block matrix (Rxy) shows only a moderate change if adjusted by the within-block correlations (i.e., Omega, see Figs 1 & 2). If the within-block correlation matrix is identity, then PLS and CCA would have identical outcomes (e.g., Fig 10).

Low within-block correlations for PLS and CCA are not preconditions for the analysis and are likely not to be the case for most data. We will discuss the use of pre-analysis PCA below, but for most real-world data, there are likely to be non-zero within-block correlations, especially when the variables within a block represent a common feature, such as brain networks in neuroimaging.

2) ***If there are high correlations in either of the within-block matrices, PLS-C and CCA will differ in terms of the singular/eigenvector weights. This can be mitigated by using the structure coefficient for CCA. However, the high within block correlations add instabilities to parameter estimates that interact with sample size in CCA.***

The second COSMO and SIMREL data showed the divergence of outcomes between PLS and CCA when the within-block correlations are high. The reason for the divergence is apparent in the comparisons of Rxy and Omega, where Omega becomes a smoother matrix (see Figs 4 &

6). After decomposition with SVD, the weights in CCA are dominated by a smaller set of variables. This could lead to the interpretation that only that smaller subset is relevant to the cross-block relationship (e.g., Fig 7). The instability of CCA has been noted more recently as well in neuroimaging [20]. The use of canonical structure coefficients can mitigate this. However, this does not mitigate instabilities in estimation with smaller sample sizes, where the high correlations interact with smaller N's, leading to near rank deficiency in the within-block correlation matrix.

3) ***In the case of very high within-block correlations, using component scores from pre-analysis PCA renders PLS-C and CCA equivalent. A pre-analysis PCA must be used with caution, however, especially if the component structure captures a known latent variable structure***

This gets back to point #1 where if the within-block correlations are low (or zero in this case) the two methods give similar results. This obviously would be the case if PCA was done on the within-block data and the component scores used for the CCA/PLS analysis. The challenge here is to be sure that the component structure is reliable. In the case where a pre-existing structure is used, such as a factor analysis for cognitive or behavioural measures or independent components analysis (ICA) for resting-state networks in fMRI [21], the user should be mindful of whether the pre-existing structure maps to their data. For example, if the factor structure is derived from a healthy population but applied to a clinical group (or more often the reverse), the factor structure may not be reliable, adding error to the subsequent analyses. This could also be the case for ICA where resting-state networks for a given group may change. The challenge then is to differentiate when the divergence of results comes from the derived CCA/PLS results per se or a lack of fit for the PCA/ICA. The component structure was dominated by a few components for the data used from the SIMREL data, meaning that reproducibility was less of a concern. However, in cases where several components are needed to capture the within-block variance, reproducibility can be compromised (see PCA section in Supplementary Material)

A second issue is more practical in that the interpretation of the PLS/CCA outcome needs to be filtered through the component loadings. For example, for the SIMREL data where pre-PCA was performed, the eigenvector weights for the X-variables would need to be interpreted in addition to which components were most dominant for a given PLS or CCA latent variable. This is an interpretive challenge but does add a level of difficulty to the analysis that is not there if the pre-analysis PCA is not used.

4) ***Null hypothesis testing should be complemented with tests for reliability and reproducibility.***

One suggestion for both PLS and CCA is that the null hypothesis testing needs to be complemented with assessing reliability and reproducibility. While statistical significance and reliability overlap, they are not redundant and complementary for a fulsome evaluation of data structures. This has been noted in early work on CCA [22]. We also noted this in early PLS work, where we used bootstrap estimation of standard errors to show situations where data were significant but not reliable and the converse where non-significant data were reliable [10]. This latter case is particularly salient when a parameter estimate is extremely low, but reliably so. This can be further extended to evaluations of stable zeros, where a parameter estimate of zero that is reliable would signal the absence of an effect, while a zero parameter with high variance would suggest noisy data.

Reproducibility is similarly vital, especially in large samples where null hypothesis testing can be somewhat indiscriminate, assigning significance to trivial effects. This was evident in COSMO set 2 and SIMREL, where LVs that were significant by permutation were not reproducible at both the global and local levels. It is noteworthy that the long-standing Bartlett's test did seem to agree with the reproducibility metrics in the full sample with the exception the first COSMO dataset.

## Final comments:

The paper was purposely focused on the generic forms of CCA and PLS. It is indeed the case that both have seen modifications to better deal with the complicated features of large data. Sparse versions of CCA, for example, are available that may be able to better deal with high dimensional data without needing a pre-analysis decomposition [23-25]. Sparse versions of PLS-R and PLS-C are also available [26, 27]. Whether these advances mitigate the concerns about reproducibility demonstrated here is beyond the scope of this paper. However, given that both generic forms are used here, the cautions and concerns of this work should be extended to derivations and related multivariate data analytic methods.

PLS and CCA reflect slightly different philosophies underlying the nature of data to be analyzed and the questions of interest. CCA comes out of the long history of GLM, wherein the focus of the model is on the unique contributions variables within a set make to the prediction of variables within another. This is reflected in the calculations for most GLM applications that adjust for redundancies in variance between variables (i.e., semi-partialling).  PLS is a more recent development that emphasizes redundancies in soft cause and effect modeling where the redundancy is kept in the analysis [28, 29]. The underlying assumption here is that redundancy is a feature of the data to be analyzed. This difference can be connected to practical examples of neuroimaging.  In the case of analysis with CCA, the outcome would emphasize the unique contributions of voxels/region to the prediction of task effects or behaviour. For PLS, the outcome would emphasize the collective contribution of sets of voxels/regions. These two approaches can be made equivalent if the redundancy amongst voxels/regions is removed or more likely concentrated into independent sets using PCA or ICA. The question of whether this is appropriate or necessary lies with the theoretical orientation of the researcher. Analytic tools are not agnostic to theory. It is incumbent upon the researcher to understand these theoretical assumptions before applying them to data.

## General Methods

Our simulations assessed the following features of CCA and PLS. All were evaluated as a function of sample size.

1) False positive rate
2) Detection of an effect
3) Reproducibility of covariances or canonical correlations (i.e., singular values S) and singular vector weights assessed both for the original population and within a given sample size.
4) The effects of analysis using PCA scores rather than the original variables.

For each simulation, the following was done. First, standard CCA and PLS-C were run on the full data and statistical assessment done both with parametric statistics for CCA and non-parametric resampling statistics (permutations and bootstrap) as described below. The analysis established the "population" ground truth.

Next random subsamples were drawn from the population 500 times, and CCA and PLS run for each subsample, along with the nonparametric statistical assessments. The sample sizes were 500, 250, 100, 50, and 20. In cases where n<p, CCA was not conducted to avoid operations on rank-deficient matrices. This was supplemented by runs on the same data where a PCA was done first to reduce the dimensionality of the data table where n<p.

**Statistical Evaluation**:

The parametric derivation of Wilks Lambda to Bartlett's test makes the statistical assessment of CCA straightforward[1]. There is no equivalent parametric derivation for PLS, so statistical evaluation was done using resampling statistics: permutation tests and bootstrap parameter estimation. We used resampling statistics for both approaches to make the outcomes more comparable.

Permutation tests were used for null hypothesis testing of the singular values. Data from one block is fixed, and the rows of the other are randomly reshuffled, breaking the original association between data blocks (i.e., resampling without replacement). For each permutation, the PLS and CCA were conducted, and singular values stored. Once 1000 permutations are complete, the original singular values for each LV are compared against their permuted counterparts to see how often a permuted singular value was higher. This count is divided by the number of permutations giving a probability for the null hypothesis test. This process is similar to that used in the original PLS-C work[9, 10]

Bootstrap resampling was used to estimate confidence intervals for the variable weights for the singular vectors for each LV.  Instead of resampling one data block without replacement, bootstrap estimation entails resampling **both** data blocks with replacement, maintaining the link between the rows of X and Y. At each iteration, PLS and CCA are performed, and singular vectors are stored. After 1000 iterations, the distributions of singular vector weights, including the original, are created, and upper and lower bounds of the 95% confidence interval for each weight defined. This is performed with the singular vector weight after multiplication by its singular value:

16) $U_s = U * S$
17) $V_s = V * S$

Since the vectors in *U* and *V* are unit length after extraction with SVD, scaling by *S* better expresses the contributions of each variable to an LV in terms of the total cross-block variance structure.

Finally, given that the sign of the weights within a singular vector is arbitrary, we adjusted for reflections on each bootstrap iteration to ensure the signs for each LV were comparable to the original. This was done using the operation:

19) $Z = U_{orig}^t * U_{boot}$

Where $U_{orig}$ is the original *U* matrix from the SVD and $U_{boot}$ is the *U* matrix from the SVD for bootstrap resampled data. Both matrices are left in their original unit length so that the matrix *Z*, contains the cosines of the angles between the vectors within the two matrices.  We examined the diagonal of Z and if the sign was negative, the vector in $U_{boot}$ was multiplied by -1 and also applied to the corresponding vector in *V*. The process thus corrects only for reflection and not rotations. We opted for this approach since much of the emphasis of our exploration was on the first LV across data sets, which is the least likely to show rotations (position changes) during resampling if the relative pattern of weights within a singular vector weight is robust.

***Detection***
Except for the first simulation with null data, detection was defined as the ratio the number of times the null hypothesis was rejected at P<0.05 over the total number of random resamplings within a given sample size. This is ***not*** a measure of statistical power.

## Reproducibility assessment

**Train-test**: This assessment focused on the magnitude of the relation between X and Y matrices, as captured by the singular value from the decomposition in eq 6 and 9.  Here we used a "*train-test*" approach where the decomposition was performed on the cross-block matrix of half the original sample, and then the singular vectors from that were applied to the cross-block correlation matrix, *M*, computed from the second half to derive their singular values (for PLS *M* is *Rxy,* and for CCA *M* is *Ω*. The assessment essentially tests whether the same LV pattern can produce as strong a relation, indicated by the derived singular value, as in the original sample.

Specifically, this is done by rearranging the terms in eq's 6 & 9 as follows:

$$20) \; M_{train} = U_{train} * S_{train} * V_{train}^t$$

Where the subscript '*train'* indicates the data from a random sample from half the observations in *X* and *Y*. To get $S_{test}$ from the derived singular vectors from the train data:

$$21) \; S_{test} = U_{train}^t * M_{test} * V_{train}$$

Where subscript "test" indicates data from remain half of the total sample (i.e., $M_{test}$). The diagonal of $S_{test}$ serves as the assessment of the strength of the relationship between *X* and *Y* in a test sample using the pattern of singular vector weights from the training sample.

The test-train resampling was performed 500 times. The mean of the distribution of the test singular values and the standard deviation was calculated, and the ratio computed akin to a z-transform. The z-test here is used as a heuristic, where the z-score would be higher if the distribution was narrow, indicating reproducibility of the mean was good.

**LV reproducibility**: The second level of reproducibility also used the same split-half resampling but calculated the similarity of singular vectors derived from each split half. Specifically, if *M1* is the cross-block matrix for a random selection of half the original sample and *M2* is the cross-block matrix for the second half, then an SVD of each of the matrices yields

$$22) \; M1 = U_1 * S_1 * V_1^t$$

And
$$23) \; M2 = U_2 * S_2 * V_2^t$$

We compute the similarity of the singular vectors:

$$24) \; U_{repro} = U_1^t * U_2$$

And

25) $V_{repro} = V_1^t * V_2$

Because singular vectors are unit length, the dot-product is the cosine similarity matrix as in eq 19. We used the absolute value for the cosines to control for arbitrary sign changes in the singular vectors during resampling.

As with the train-test singular value estimates, the split-half resampling was done 500 times and the mean of $U_{repro}$ and $V_{repro}$ distributions converted to a z-score interpreted the same as for the distribution of singular values in the train-test procedure described above.

We also validated the use of the z-score test for reproducibility by performing the assessment on a null distribution generated using permutation resampling as described above. In all cases, and for all datasets, the z-scores for these null tests for singular value and singular vectors seldom exceeded 1.65 (see example in supplementary material section S1).

All analyses were performed in Matlab R2018a.

**Declaration**:
Ethics approval and consent: Not applicable

Consent for publication: Not applicable

Availability of data and materials: The code used for analyses and the data converted into Matlab matrices are available via our lab's Github repository https://github.com/McIntosh-Lab-RRI/PLS-CCA

Competing Interests: there are no competing interests

Authors' contributions: sole authored paper

Funding: This work is supported by the Natural Sciences and Engineering Council of Canada Grant RGPIN-2018-04457 and Canadian Institutes of Health Research Grant PJT 168980.

Acknowledgements: Access to the COSMO survey data presented here was provided generously by Lisa Felgendreff and Cornelia Betsch. I am grateful for the discussions with Herve Abdi and Viktor Jirsa about this work, and Sarah Faber for helping to proof the paper.

# Comparison of Canonical Correlation and Partial Least Squares analyses of simulated and empirical data.

## Supplementary Material

### S1. Reproducibility Assessments

We compared the train-test and split-half assessments on data using a permuted version of one block, which essentially creates a null model to compare with the unpermuted (original) data. We performed this null assessment for all data sets and the z-scores for the null models seldom were above 1.96. However, in cases where it does, the null model provides an idea of expected value of the reproducibility metrics under the null. This is depicted in the two examples below. These data are from the PLS analysis of the second COSMO sample, which has two strong LVs and two that are unreliable. The z-values for the reproducibility distributions show this, where both the singular values for train-test and singular vectors from split-half (for matrix X only in this example). The pattern was essentially the same for CCA.

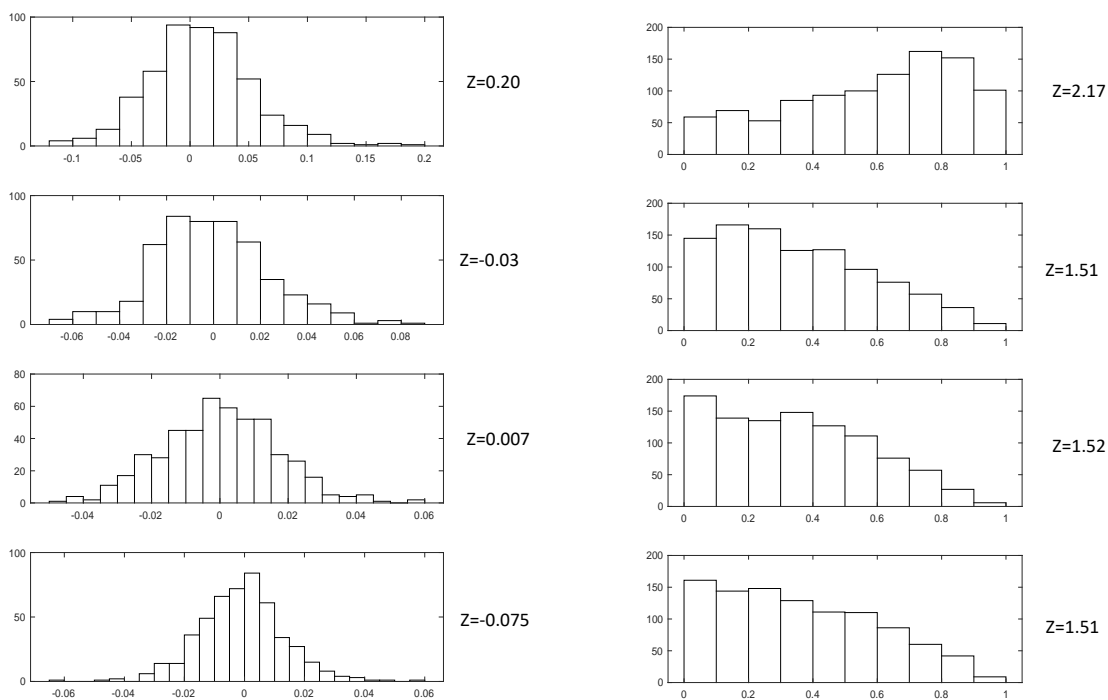

*Figure S1. Distributions for reproducibility metrics for singular values (left) and the X singular vector (U). Z-values are the ratio of the distribution mean over its standard deviation. The data*

*come from the second COSMO dataset.  Each row corresponds to a Latent Variable from the PLS analysis.*

For the null model, the distributions had substantially lower z-values for the first two LVs, especially for the singular values. The utility of the null model is shown for singular vector for the first LV - 2.17 – which by some criteria would be considered "significant". This expected value under the null can be compared with the obtained value, which in this case is tremendously larger at 933.5. However, the z-values for the lower order LVs in both the null and original model are essentially the same (~1.5) confirming the lack of reliability for these LVs

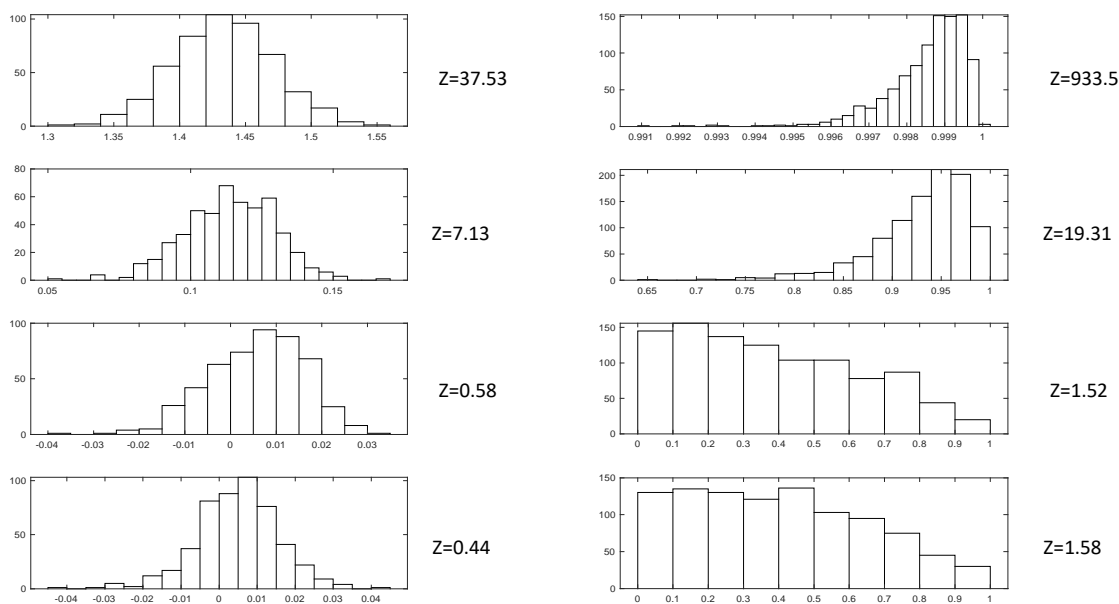

*Figure S2. Distributions for reproducibility metrics for singular values (left) and the X singular vector (U) for permuted data to give a null distribution. The data come from the second COSMO dataset, where the rows for the Y matrix were permuted for iteration. Z-values are the ratio of the distribution mean over its standard deviation. Each row corresponds to a Latent Variable from the PLS analysis.*

## S2. PCA stability

Using the PCA scores as input to PLS and CCA is a common practice, but is not without its issues. In the situation where the components are meant to reflect an underlying latent structure in the population, the stability of the component weights with the given sample needs to be evaluated. In addition, the interpretation of the PLS or CCA output requires the secondary step where the contributions of the original variables are filtered through the eigenvectors and then the singular vectors from CCA/PLS.

For the SIMREL data (Data Set 4), the eigenspectrum had a steep drop (Fig S3), indicating that most of the variance is captured in the first few components. The steep drop also suggests that the component structure would likely be preserved across different resampling regimes, especially with smaller sample sizes.

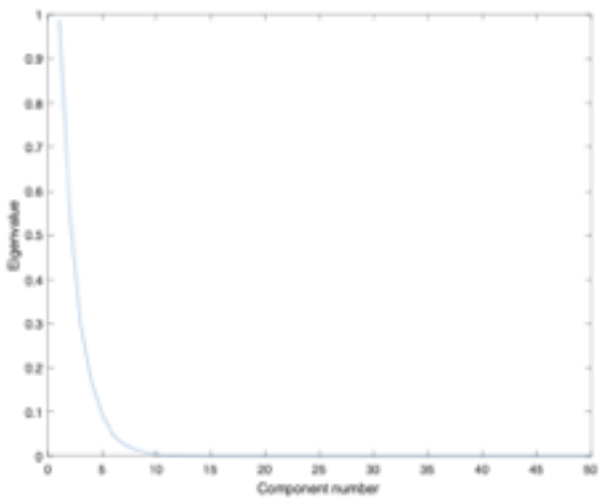

*Figure S3. Eigenspectrum for SIMREL data.*

This is indeed the case where the cosine of the first seven eigenvectors from the full sample with the eigenvectors from the smaller sample sizes all showed strong reproducibility as indicated by the z-score for the cosine distribution (Table S1). Arbitrary reflections of the eigenvector can affect the similarity we test reproducibility with and without correction (z-score for distribution without correct are in parenthesis in the tables). For all sample sizes, the first and second PCs were reproduced well regardless of correction. Without correction, the z-score for PC2 dropped to near zero for all sample sizes.

Table S1. z-score value from distribution of cosine angle between population components and subsamples.  Values in parentheses from distribution with no correction for eigenvector reflections

| Sample Size | PC1 - z-score | PC2 - z-score |
|---|---|---|
| 500 | 234.6 (234.6) | 179.75 (0.138) |

| 250 | 110.5 (110.5) | 87.7 (0.02) |
| 100 | 23.8 (23.8) | 20.5 (0.01) |
| 50 | 9.9 (9.9) | 8.1 (-0.03) |
| 20 | 4.7 (4.4) | 3.2 (0.07 |

In a situation where the eigenspectrum is flatter, the reproducibility drops. Here we generate a dataset with 15 X variables with SIMREL, with a more gradual change in the eigenvalue spectrum (Figure S4).

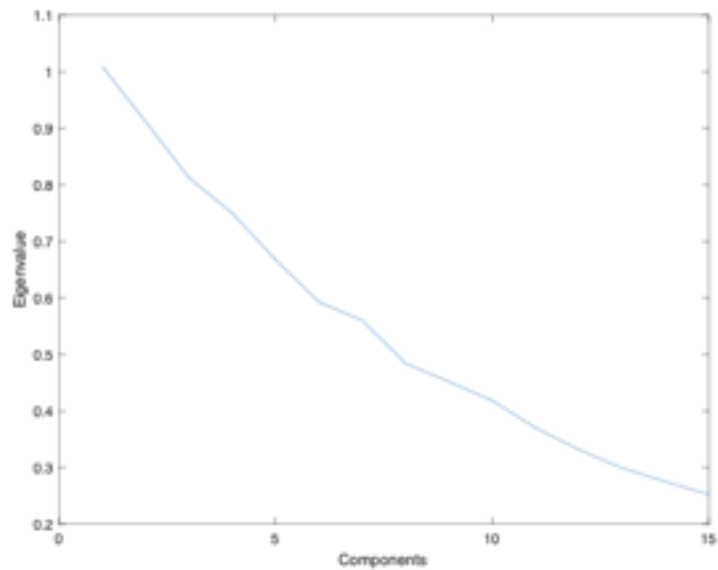

*Figure S4. Eigenspectrum for SIMREL data with variance spread across more componoents.*

The reproducibility of the PCA solution relative to the full sample was extremely low without correction for reflections. Even with such a correction, the z-score for reproducibility was only high for N=500 and N=250, and only for the first two components.

Table S2. z-score value from distribution of cosine angle between population components and subsamples. Values in parentheses from distribution with no correction for eigenvector reflections

| Sample Size | PC1 - z-score | PC2 - z-score |
| --- | --- | --- |
| 500 | 4.5 (0.28) | 3.1 (-0.18) |
| 250 | 3.3 (0.25) | 2.2 (-0.24) |
| 100 | 2.4 (0.19) | 1.73 (-0.15) |

| 50 | 2.02 (0.19) | 1.57 (-0.23) |
| 20 | 1.8 (0.07) | 1.5 (-0.11) |